\documentclass[lettersize,journal]{IEEEtran}
\usepackage{amsmath,amsfonts}
\pdfoutput=1
\usepackage{array}
\usepackage[caption=false,font=normalsize,labelfont=sf,textfont=sf]{subfig}
\usepackage{textcomp}
\usepackage{stfloats}
\usepackage{url}
\usepackage{verbatim}
\usepackage{graphicx}
\usepackage{cite}
\usepackage{booktabs}
\usepackage[justification=centering]{caption}
\usepackage{amssymb}
\usepackage{bm}
\usepackage[ruled,linesnumbered]{algorithm2e}
\usepackage[colorlinks,linkcolor=blue,anchorcolor=blue,citecolor=blue,bookmarks=blue]{hyperref}
\begin{document}

\title{Joint Optimization of Data- and Model-Driven Probing Beams and Beam Predictor}

\author{Tianheng~Lu, Fan~Meng, Zhilei~Zhang, Yongming~Huang,~\IEEEmembership{Senior Member,~IEEE},\\ Cheng~Zhang,~\IEEEmembership{Member,~IEEE}, Xiaoyu~Bai
% <-this % stops a space
\thanks{This work was supported in part by the National Key R\&D Program of China under Grant 2020YFB1806600 and the National Natural Science Foundation of China under Grant No. 62225107, 62001103 and 62201394, and the Fundamental Research Funds for the Central Universities under Grant 2242022k60002.}
\thanks{T. Lu, F. Meng, Z. Zhang, X. Bai, C. Zhang and Y. Huang are with the Purple Mountain Laboratories, Nanjing 211111, China (e-mail: th\_lu@seu.edu.cn; mengfan@pmlabs.com.cn; zhangzhilei@pmlabs.com.cn; baixiaoyu@pmlabs.com.cn; zhangcheng\_seu@seu.edu.cn; huangym@seu.edu.cn). C. Zhang and Y. Huang are (also) with the National Mobile Communications Research Laboratory, School of Information Science and Engineering, Southeast University, Nanjing 210096, China.}% <-this % stops a space
}

\markboth{Journal of \LaTeX\ Class Files,~Vol.~14, No.~8, August~2021}%
{Shell \MakeLowercase{\textit{et al.}}: A Sample Article Using IEEEtran.cls for IEEE Journals}


\maketitle

\begin{abstract}
	
Hierarchical search in millimeter-wave (mmWave) communications incurs significant beam training overhead and delay, especially in a dynamic environment. Deep learning-enabled beam prediction is promising to significantly mitigate the overhead and delay, efficiently utilizing the site-specific channel prior. In this work, we propose to jointly optimize a data- and model-driven probe beam module and a cascaded data-driven beam predictor, with limitations in that the probe and communicate beams are restricted within the manifold space of uniform planer array and quantization of the phase modulator. First, The probe beam module senses the mmWave channel with a complex-valued neural network and outputs the counterpart RSRPs of probe beams. Second, the beam predictor estimates the RSRPs in the entire beamspace to minimize the prediction cross entropy and selects the optimal beam with the maximum RSRP value for data transmission. Additionally, we propose to add noise to the phase variables in the probe beam module, against quantization error. Simulation results show the effectiveness of our proposed scheme.

\end{abstract}


\section{Introduction}

\IEEEPARstart{W}{ith} sufficient bandwidth and potentially high data rates in B5G/6G communications~\cite{7959169}, millimeter-wave (mmWave) communication technology has become a hot topic of research~\cite{Li:2018wp}. High-frequency signals suffer significant attenuation in propagation, and large-scale antenna arrays with beamforming is introduced to compensate for the path loss and simultaneously improve anti-interference capability~\cite{7010533}. However, the traditional hierarchical beam alignment/tracking (BA/T) incurs a large training overhead, resulting in inefficient beam training. Therefore, a low overhead and stable beam training method should be proposed urgently.

Conventional model-driven beam alignment schemes include exhaustive and hierarchical searches~\cite{6600706}, which are unable to utilize the a priori knowledge of the channel state information (CSI) and have drawbacks such as high overhead and error propagation. In contrast, deep learning-based schemes can effectively extract the CSI prior in temporal, frequency, and spatial domains to improve the prediction performance~\cite{9954418,9598911,9269463,9944873}.

Many studies have focused on the design of beam predictors, e.g., \cite{9129762, 9690703}, and their idea can be summarized as using deep neural networks to find the mapping of a certain measured quantity to the optimal beam. Compared to traditional schemes, they improve the accuracy while reducing the interpretability and generalization of the model. The researchers in \cite{7248501} jointly consider beam width design and power allocation strategy, but this scheme is difficult to obtain the global optimal solution, directly. Reference \cite{9690703} learns a set of probe codebooks for a specific scenario, and it can be seen through simulation that the learned probe codebooks perform better than the wide beam. However, the method has more training parameters, especially when the ULA antenna is extended into a UPA antenna.

In this work, we predict the optimal beam in beamspace with RSRPs of a small number of probe beams. The principle of beam prediction is to utilize the airspace beam correlation to realize nonlinear interpolation, and the performance of beam prediction is mainly affected by two aspects: the probe beams and the beam predictor. We take beam prediction performance as optimization objective, and the probe beams and the beam predictor as the optimization variables, to achieve low beam training overhead and approximate the optimal intelligent real-time BA/T performance. The main contributions are summarized as follows.

\subsubsection{Data- and model-driven Probing Beam Training}

To compensate for the lack of physical understanding and poor generalization ability of traditional pure data-driven schemes, we propose a complex-value neural network (CVNN) that employs DFT-like manifold to generate probe beams in the training process. CVNN has the advantages of fewer training parameters and better generalization ability, which can effectively extract the features of the mmWave propagation environment and empower the downstream prediction task.

\subsubsection{Beam Domain Equivalent Variables}

We propose to train the CVNN with equivalent variables of the horizontal and vertical angles, i.e., the variables in beamspace. The angle-based beams only cover a small beamspace and the counterpart gradient is not smooth w.r.t. the angle variables. While, the beam-based variables cover the entire beamspace and have a smooth gradient, leading to better beam prediction accuracy.

\subsubsection{Noise Adding Technique}

Considering the limited phase resolution of practical mmWave devices, the learned probe beams have significant performance degradation after phase quantization. To address this issue, we propose additional noise on the phases of probe beams during training, to simulate quantization operation.

\textit{Notations}: Lower-case and upper-case boldface letters $\mathbf{a}$ and $\mathbf{A}$ denote a vector and a matrix, respectively; $\mathbf{A}^{\mathsf{H}}$ and $\mathbf{A}^{\mathsf{T}}$ denote the conjugate transpose and transpose of matrix $\mathbf{A}$; $ |\cdot| $, $ \otimes $ respectively denote absolute and Kronecker product operators. $ \mathbb{E}\{\cdot\} $, $ \mathbb{R} $, $ \mathbb{C} $ represent the expectation, real and complex fields.

\section{System Model and Problem Formulation}\label{sec:system}

\subsection{System Model}

%The scheme can be easily extended to cellular or cell-free networks with multiple BSs, RFs and MUs via multiple access.

Consider a link-level downlink mmWave multiple-input single-output (MISO) communication system consisting of a single base station (BS) and a mobile user (MU). The BS is equipped with a large uniform planar array (UPA) where $N$ antennas are connected to a radio frequency (RF) chain, and the MU has an isotropic antenna. The BS uses the codewords in a DFT codebook $ \mathcal{A} = \{{\bf a}_i\}_{i=1}^{N} $ where $ {\bf a}_i \in \mathbb{C}^{N\times 1}, \forall i $, to probe the channel and communicate with the MU. Based on the 3GPP channel modeling, the downlink channel $\mathbf{h} \in \mathbb{C}^{N\times 1}$ is characterized as a superposition of $ M $-paths propagation due to interactions (reflections, diffractions, penetrations, scattering) at stationary obstacles(hills, buildings, towers) and mobile objects(cars, pedestrians), given as
\begin{equation}
\mathbf{h}=\sum_{m=1}^{M} \alpha_{m} \mathbf{\psi} (\phi_{m},\theta_{m}),
\end{equation}
where $\alpha$ is the complex gain coefficient, $\phi$ and $\theta$ respectively are the horizontal and vertical angles, and the UPA response $\psi$ is expressed as
\begin{equation}
\label{equ:array_response}
\boldsymbol{\psi}(\phi, \theta) = \boldsymbol{a}_{\textup{xy}}(\phi, \theta) \otimes \boldsymbol{a}_{\textup{z}}(\theta),
\end{equation}
where
\begin{align}
\boldsymbol{a}_{\textup{xy}}(\phi, \theta) & = \frac{1}{\sqrt{N_{\phi}}}[1, e^{\jmath\pi\sin\phi\sin\theta}, \cdots, e^{\jmath \pi (N_{\phi}-1) \sin\phi\sin\theta}]^{\mathsf{T}},\\
\boldsymbol{a}_{\textup{z}}(\theta) & = \frac{1}{\sqrt{N_{\theta}}}[1, e^{\jmath \pi\cos\theta}, \cdots, e^{\jmath \pi (N_{\theta}-1) \cos\theta}]^{\mathsf{T}},
\end{align}
where $N_{\phi}$ and $N_{\theta}$ represent the number of antennas in the horizontal and vertical directions of the array, respectively, and $N=N_{\theta}\times N_{\phi} $ is the total number of antennas.
\subsection{Problem Formulation}

In a beam alignment process, the BS first transmits beams in a probe beam codebook to sense the downlink channel. The probe beam codebook $ \mathbf{W} $ is composed of $L$ probe beams, and $ \mathbf{W} =[\mathbf{w}_1,\cdots,\mathbf{w}_L]$ where $ \mathbf{w}_l \in \mathcal{A}, \forall l \in \{1,\cdots,L\} = \mathcal{L} $. Then, the MU receives the probe signals and feedbacks the counterpart RSRPs to the BS. It has to be noticed that the transmitting and feedback should all be finished in the same channel coherence time. The $ l $-the entry of the MU's feedback RSRPs $ \mathbf{z}=[z_{1},\dots,z_{L}]^{\mathsf{T}} $ is written as
\begin{equation}\label{equ:x_pi}
z_{l} = \sqrt{P} \mathbf{h}^{\mathsf{H}}\mathbf{w}_{l}s+n_{l}, \forall l \in \mathcal{L},
\end{equation}
where $P$ is BS transmit power, $s$ is baseband signal with unit power, and $n_{l}$ is additive white gaussian noise. We denote receive signal power as $ \mathbf{x} = |\mathbf{z}|^2 $, and $ \mathbf{x} $ is further quantized as RSRPs $ \mathbf{x}_{q} $ by a map $ g $. Using RSRPs $ \mathbf{x}_{q} $ as input, the BS learns to infer the optimal transmit beam in the DFT codebook by a learnable map $f: \mathbf{x}_{q} \rightarrow i^{*}$. Moreover, the probe beam codebook is parameterized by learnable parameters $\{\phi_l,\theta_l\}_{l=1}^L$. Finally, the joint optimization problem of data- and model-driven probing beams and beam predictor is expressed as follows
\begin{subequations}\label{equ:problem}
\begin{align}
\min_{\bm{\theta}, \bm{\phi}, f}& ~\mathop{\mathbb{E}}_{\mathbf{h},\mathbf{n}} \left\{d(\hat{\mathbf{y}}, \mathbf{y}^{\textup{tar}})\right\}\\
% s.t.& ~ |\mathbf{z}|=|\sqrt{P_{T}} \mathbf{h}^{\mathsf{H}}\mathbf{W}s+\mathbf{n}|,\\
& ~ \mathbf{W}=\psi(\bm{\theta}, \bm{\phi}),\\
& ~ \theta_{j} \in [0, \pi],\forall l \in \mathcal{L},\\
& ~ \phi_{j} \in [-\frac{\pi}{2},+\frac{\pi}{2}],\forall l \in \mathcal{L},\\
& ~ \eqref{equ:x_pi}, \nonumber\\
& ~ \mathbf{x}_{q}=g(|\mathbf{z}|^2),\\
& ~ \hat{\mathbf{y}}=f(\mathbf{x}_{q}),\\
& ~ \hat{i}^{*} = \arg\max \hat{\mathbf{y}},
\end{align}
\end{subequations}
where $d$ is a distance function, $ \mathbf{y}^{\textup{tar}} $ is the target of the beam predictor. Map $g$ converts the received signal into RSRP, i.e., $g: \max(\min(10\log_{10}(\cdot), -40), -140) $, indicating the RSRP is capped between $-140$ to $-40\,$dBm. Particularly, the RSRP is not quantized during training and is quantized with 1 dBm resolution for online inference.

\section{Data- and Model-Driven Solution of Probe Beam Training and Beam Predictor}\label{sec:algorithm}

\begin{figure}[h]
\centering
\includegraphics[width=3.2in]{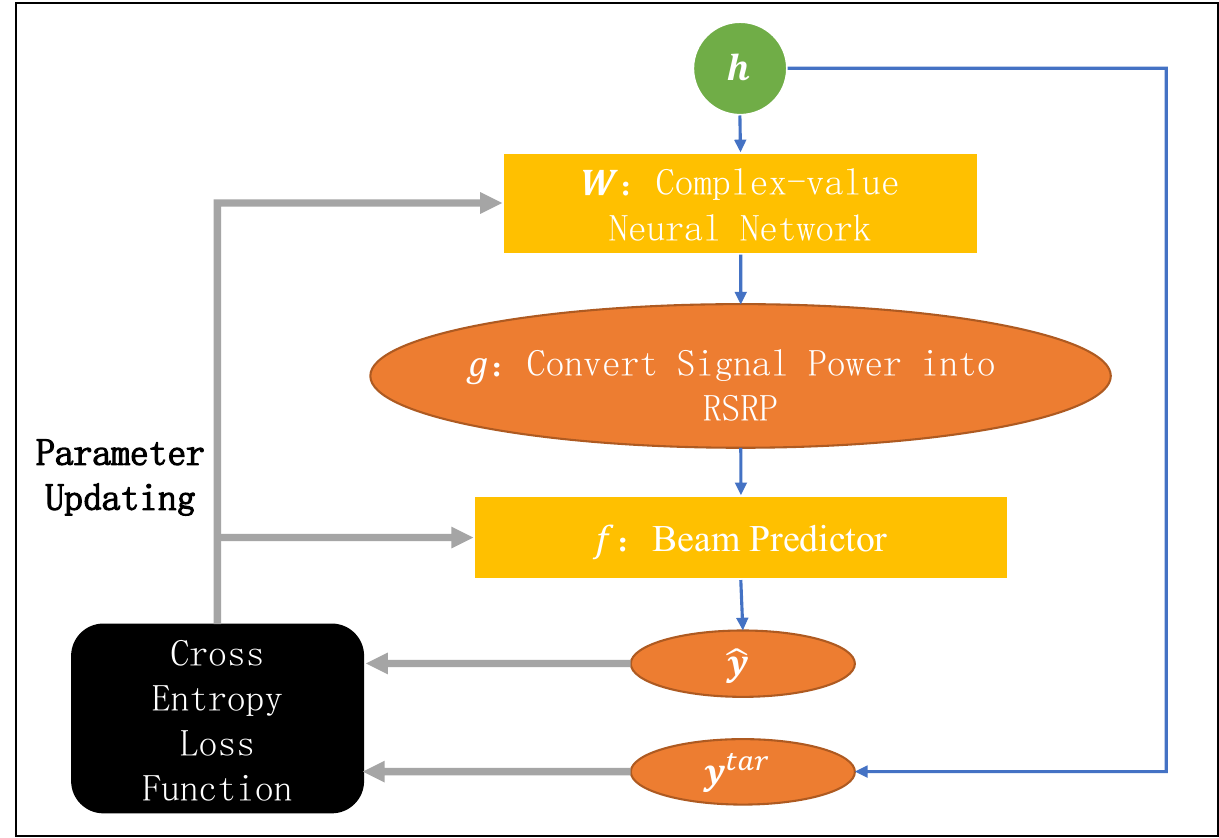}
\caption{The illustrative procedure of the proposed scheme.}
\label{fig:scheme}
\end{figure}

As shown in Fig.~\ref{fig:scheme}, the overall learning problem \eqref{equ:problem} is composed of a probe beam training module with $\bm{\theta}, \bm{\phi}$, and a cascaded beam predictor with $f$, and we discuss the counterpart solution. In general, the probe codebook and the beam predictor are parameterized with separate neural networks and jointly trained in an end-to-end manner, via the stochastic gradient descent method. In this way, the probe beams are indirectly optimized to assist in the downstream prediction effort.

\subsection{Probe Beam Training}

We propose to design the probe beam training module with a complex-valued neural network(CVNN), and derive the output gradient w.r.t. the learnable parameters $ \bm{\theta}, \bm{\phi}$.

To directly characterize the energy magnitude and at the same time facilitate the computation, we take $\mathbf{z}=[(x_{1}^{r})^{2}+(x_{1}^{i})^{2},\cdots,(x_{L}^{r})^{2}+(x_{L}^{i})^{2}]^{\mathsf{T}}$ as an input to the subsequent multi-classifier. Denote the loss function as $J$, according to the derived chain rule, the partial gradient w.r.t. $\phi
_{j}$ is expressed as
\begin{equation}
\frac{\partial J}{\partial\phi_{j}}=\frac{\partial J}{\partial|z_{j}|^{2}}\frac{\partial|z_{j}|^{2}}{\partial z_{j}}\frac{\partial z_{j}}{\partial\phi_{j}},
\end{equation}
where $\frac{\partial J}{\partial|z_{j}|^{2}}$ can be derived by automatic differentiation with Pytorch in implementation. Although $ |z_{j}|^{2}$ is not complex differentiable w.r.t. $z_{j}$, its gradient can be computed by treating the real and imaginary parts of $z_{j}$ separately: $\frac{\partial|z_{j}|^{2}}{\partial z_{j}}=[2z_{j}^{r},2z_{j}^{i}].$ And the expression of $\frac{\partial z_{j}}{\partial\phi_{j}}$ is written as
\begin{equation}
\frac{\partial z_{j}}{\partial\phi_{j}}= \begin{bmatrix} \frac{\partial z_{j}^{r}}{\partial\phi_{j}} \\ \frac{\partial z_{j}^{i}}{\partial\phi_{j}} \end{bmatrix}=
\begin{bmatrix}
\sum_{n=1}^{N}(\frac{\partial w_{jn}^{r}}{\partial\phi_{j}}h_{n}^{r}-\frac{\partial w_{jn}^{i}}{\partial\phi_{j}}h_{n}^{i}) \\
\sum_{n=1}^{N}(\frac{\partial w_{jn}^{r}}{\partial\phi_{j}}h_{n}^{i}+\frac{\partial w_{jn}^{i}}{\partial\phi_{j}}h_{n}^{r})
\end{bmatrix},
\end{equation}
where
\begin{subequations}
\begin{align}
\frac{\partial w_{jn}^{r}}{\partial\phi_{j}}=&\frac{1}{\sqrt{N}}\{-\sin\{\pi [(\lceil\frac{n}{N_{\theta}}\rceil-1)\sin\phi_{j}\sin\theta_{j} \nonumber \\
&+(n-\lceil\frac{n}{N_{\theta}}\rceil N_{\theta}+N_{\theta}-1) \cos\theta_{j}]\}\}\nonumber \\
&\cdot\pi (\lceil\frac{n}{N_{\theta}}\rceil-1)\sin\theta_{j}\cos\phi_{j},\\
\frac{\partial w_{jn}^{i}}{\partial\phi_{j}}=&\frac{1}{\sqrt{N}}\{\cos\{\pi [(\lceil\frac{n}{N_{\theta}}\rceil-1)\sin\phi_{j}\sin\theta_{j} \nonumber \\
&+(n-\lceil\frac{n}{N_{\theta}}\rceil N_{\theta}+N_{\theta}-1) \cos\theta_{j}]\}\} \nonumber \\
&\cdot\pi (\lceil\frac{n}{N_{\theta}}\rceil-1)\sin\theta_{j}\cos\phi_{j}.
\end{align}
\end{subequations}\par
Similarly, for the update of $\theta_{j}$,
\begin{equation}
\frac{z_{j}}{\partial\theta_{j}}= \begin{bmatrix} \frac{z_{j}^{r}}{\partial\theta_{j}} \\ \frac{z_{j}^{i}}{\partial\theta_{j}} \end{bmatrix}=
\begin{bmatrix}
\sum_{n=1}^{N}(\frac{\partial w_{jn}^{r}}{\partial\theta_{j}}h_{n}^{r}-\frac{\partial w_{jn}^{i}}{\partial\theta_{j}}h_{n}^{i}) \\
\sum_{n=1}^{N}(\frac{\partial w_{jn}^{r}}{\partial\theta_{j}}h_{n}^{i}+\frac{\partial w_{jn}^{i}}{\partial\theta_{j}}h_{n}^{r})
\end{bmatrix},
\end{equation}
where
\begin{subequations}
\begin{align}
\frac{\partial w_{jn}^{r}}{\partial\theta_{j}}=\frac{1}{\sqrt{N}}\{&-\sin\{\pi [(\lceil\frac{n}{N_{\theta}}\rceil-1)\sin\phi_{j}\sin\theta_{j} \nonumber \\
&+(n-\lceil\frac{n}{N_{\theta}}\rceil N_{\theta}+N_{\theta}-1) \cos\theta_{j}]\}\} \nonumber \\
\cdot \pi [(\lceil&\frac{n}{N_{\theta}}\rceil-1)\sin\phi_{j}\cos\theta_{j} \nonumber \\
&-(n-\lceil\frac{n}{N_{\theta}}\rceil N_{\theta}+N_{\theta}-1) \sin\theta_{j}], \\
\frac{\partial w_{jn}^{r}}{\partial\theta_{j}}=\frac{1}{\sqrt{N}}&\{\cos\{\pi [(\lceil\frac{n}{N_{\theta}}\rceil-1)\sin\phi_{j}\sin\theta_{j} \nonumber \\
&+(n-\lceil\frac{n}{N_{\theta}}\rceil N_{\theta}+N_{\theta}-1) \cos\theta_{j}]\}\} \nonumber \\
\cdot \pi [(\lceil&\frac{n}{N_{\theta}}\rceil-1)\sin\phi_{j}\cos\theta_{j} \nonumber \\
&-(n-\lceil\frac{n}{N_{\theta}}\rceil N_{\theta}+N_{\theta}-1) \sin\theta_{j}].
\end{align}
\end{subequations}
After offline training, the parameters of the probe beams are further quantized for online deployment.

\subsection{Beam Predictor}

The beam predictor infers the beam with maximum RSRP value by extracting the prior information implicitly embedded in the CSI dataset that reflects the site-specific propagation environment. After the CVNN outputs the power $\mathbf{z}$ of the probe beams and quantized by the map $ g $, the RSRPs $ {\bf x}_q $ is fed into the beam predictor which is enabled by a deep neural network (DNN). The DNN output is the distribution of the optimal beam in the entire DFT space, i.e., $\hat{\bf y} \in \mathbb{R}^{N\times1}$. The optimal beam, i.e., $i^* = \arg \max \hat{\bf y} $, is selected as the communication beam. We use cross entropy as the distance function, so the activation function of the output layer is softmax: $\sigma(z_i) = \frac{e^{z_{i}}}{\sum_{j=1}^N e^{z_{j}}}, i=\{1,\dots,N\}$, and the target label $ {\bf y}^{\textup{tar}} $ is a one-hot vector.

Through the output layer, we can either directly get the predicted optimal beam, or for the consideration of robustness, re-probe the Top-$ K $ beams with the largest probabilities. Then the beam with the highest RSRP is regarded as the optimal beam, whose procedure is similar to the two-level search in 3GPP, but the searching space is greatly reduced with our proposed scheme.

\subsection{Probe Beam Training with Beamspace Variables}

At the initial stage of the study, our scheme treats the horizontal angle $\phi$ and vertical angle $\theta$ of the probe beam as a set of trainable parameters, to generate DFT-like beams. However, the beam prediction accuracy is not significantly improved. In fact, the angle-based representation of beams is restricted in the range $[-60, 60]^{\circ}$, and the beam width is angle-related. For example, the beam around $0^{\circ}$ is thinner than the beam around $ \pm 60^{\circ}$. This indicates the loss gradient is not smooth w.r.t. horizontal-vertical angles, resulting in difficulties during training.

To enlarge the probe range, we propose to use the equivalent variables $u = \sin\phi\sin\theta$ and $v = \cos\theta$ in the beam domain (both in range $[-1, 1]$) instead of the angle domain (in range $[-\pi/2, \pi/2]$), to design the learnable probe beams. Moreover, the loss gradient is uniform w.r.t. the beam-domain variables $u, v$.

\subsection{Phase Quantization}\label{subsec:quan}

In practical deployment, the analog precoders usually have limited phase resolution $B$, such as $3-7$ bits. During training, the effects of phase quantization are not considered, since back-propagation can be ruined by the quantization operations. However, the learned model can be not robust to the quantizations, to simulate the effects of phase quantization. To address this issue, we propose to add noise to the phases during training. Particularly, the additional noise follows uniform distribution in range $[-1/2^B, 1/2^B]^{\circ}$.

%\begin{algorithm}[h]
%\caption{Model-and-Data-Driven Beam Predictor(offline training)}
%\label{alg:alg_training}
%\KwIn{Dataset $ \mathcal{D} $, maximal probing beam number $ L $.}
%\KwOut{Learned predictor $ f $, probe codebook $W$.}
%\BlankLine
%Initialize
%
%\For{$ i = 1 $ to $ epochs $}{
%\For{ $ j = 1 $ to $ batch\_num $}{
%Deliver channel statement vector $h$ to CVNN.
%CVNN outputs the power of the detected signal $\mathbf{z}$, which is then converted into RSRP.
%The probability distribution of the optimal beam is given by the beam predictor.
%Perform gradient descent and update parameters $\sin\phi\sin\theta$ and $\cos\theta$.
%}
%}
%\end{algorithm}

\begin{algorithm}[h]
\caption{Data-Driven Beam Predictor (online inference)}
\label{alg:alg_inference}
{\bf Initialize}: learned probe codebook $\bf W$ after quantization, learned beam predictor $ f $;

\KwOut{communication beam ${\bf w}_{i^{*}}$.}
\BlankLine

The BS senses the downlink channel with $\bf W$;

The UE feedbacks the counterpart RSRPs to the predictor $f$ at the BS side;

The BS selects and re-probes the top-$ 5 $ beams, then gets the counterpart RSRPs;

The BS selects the beam with maximum RSRP for data transmission, i.e., ${\bf w}_{i^{*}}$.

\end{algorithm}

In summary, the data- and model-driven probe beam training and beam predictor takes channel $\mathbf{h}$ as inputs and optimizes learnable probe beams during the training phase, whose mathematical form naturally has DFT-like manifold. In the deployment phase, the parameters of the probe beams are further quantized by the phase resolution of analog devices. The online inference process is demonstrated in Algorithm~\ref{alg:alg_inference}.

%offline joint training process is described in Algorithm~\ref{alg:alg_training}, and the 

%The trainable parameters of the CVNN are initially set to be the horizontal angle $\phi$ and vertical angle $\theta$ of the probe beam array, but after subsequent experiments, it can be found that using their corresponding trigonometric values, i.e., $ \sin\phi\sin\theta$ and $ \cos\theta$ , as the training parameters can effectively improve the model performance, which will be discussed later.

\section{Simulations}\label{sec:sim}

\subsection{Configurations}

%\par
%Consider a single cell scenario with a set of BS and MU in the cell, the BS is equipped with $16\times 8$ UPA antenna arrays and the MU possesses a single antenna. The number of probe beams is 8, and the BSs are built-in with a $128\times 128$ DFT-codebooks.

To evaluate the performance of our proposed data- and model-driven scheme, the mmWave channel is established as a map-based deterministic model according to 3GPP 38.901~\cite{3GPP}, and stochastic clusters are also introduced. The DNN-based beam predictor consists of one input layer, three hidden layers, and one output layer. Scenario-related details and the specific configuration of the DNN are summarized in Table \ref{tab:sim_config} and Table \ref{tab:dnn_config}, respectively.

\begin{table}[htb]
\centering
\setlength{\tabcolsep}{0.5mm}{
\small
\centering
\caption{Simulation Configurations of Scenario}
\begin{tabular}{c|c}
\toprule
Name & Value\\
\midrule
%BS number & $1$ \\
%MU number & $1$\\
BS antenna number & $ 16 \times 8 $\\
MU antenna number & $ 1 $ \\
Carrier frequency $ f_{c} $ & $ 30\; \textup{GHz} $\\
Bandwidth $ B $ & $ 100\; \textup{MHz} $ \\
noise power spectral density & $ -174 \; \textup{dBm/Hz} $\\
probing beam number $ L $ & $ 8 $\\
symbol duration $T_{s}$ & $8.92 \; \mu\textup{s}$\\
time-slot duration $T_{c}$ & $20 \; \textup{ms}$\\
\bottomrule
\end{tabular}
\label{tab:sim_config}}
\end{table}

\begin{table}[htb]
\centering
\setlength{\tabcolsep}{0.5mm}{
\small
\centering
\caption{Configurations of DNN}
\begin{tabular}{c|c}
\toprule
Name & Value\\
\midrule
Input layer nodes number & $8$ \\
Hidden layer 1 nodes number & $ 200 $\\
Activation function 1 & LeakyReLU(0.04)\\
Hidden layer 2 nodes number & $ 200 $ \\
Activation function 2 & LeakyReLU(0.04)\\
Hidden layer 3 nodes number & $ 200 $ \\
Activation function 3 & ReLU \\
Output layer nodes number & $ 128 $\\
\bottomrule
\end{tabular}
\label{tab:dnn_config}}
\end{table}

For performance validation, two metrics are considered: 1) Top-$ K $ beam prediction accuracy, the probability that the optimal beam appears in the top $ K $ predicted beams (sorted by predicted RSRP). 2) Effective achievable rate (EAR), which is defined as
\begin{equation}
\textup{EAR} \triangleq \mathbb{E}_{{\bf h}, {\bf n}_{x}}\left\{\left(1 - \frac{L T_{\textup{s}}}{T_{\textup{c}}}\right) \log_2 \left(1 + \frac{|{\bf A}[:, i^{\ast}]{\bf h}|^2}{\sigma_x^2}\right)\right\},
\end{equation}
where $ T_{\textup{s}} $ and $ T_{\textup{c}} $ respectively are the durations of a symbol and a time slot.

The $baseline$ scheme uniformly chooses probe beams from the DFT codebook on the beamspace. The reference scheme~\cite{9690703} treats all elements in $ \bf W $ as learnable parameters with constant modulus constraint. The whole end-to-end network uses the cross-entropy function as the loss function and is trained for 200 epochs using the Adam optimizer on the simulation platforms: Python 3.8.13, Pytorch 1.10.2.

\subsection{Training Convergence Speed}

%In addition to the metrics mentioned above, considering the short coherence time of the mmWave channel, beam training is highly time-sensitive, and a good beam training scheme should be able to achieve high accuracy in a short period. 

\begin{figure}[h]
\centering
\includegraphics[width=3in]{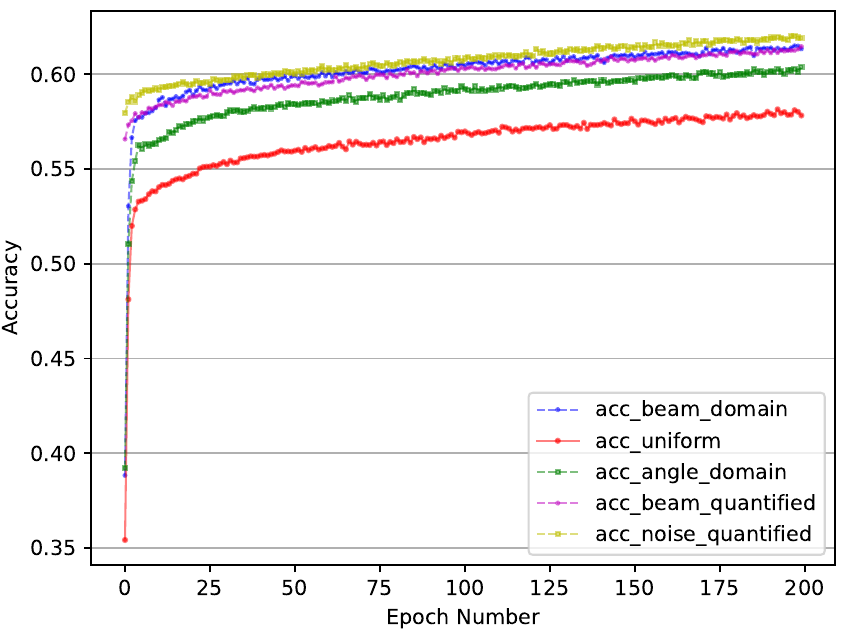}
\caption{The top-3 prediction accuracy versus training epoch.}
\label{fig:speed}
\end{figure}

This subsection focuses on the training convergence speed. In Fig.~\ref{fig:speed}, the uniform probe beam scheme performs poorly, and even after 200 epochs of training, it only obtains an accuracy rate equivalent to 10 epochs of the proposed schemes. While the training speed of our proposed schemes is faster, the beam domain scheme has an obvious speed gain compared with the angle domain scheme. In addition, when quantization noise is taken into account, the training speed of the proposed scheme can be greatly improved  further.

\subsection{Top-$ K $ Beam Prediction Accuracy}

After training, the results evaluated by Top-$ K $ beam prediction accuracy are investigated, where $K \in \{1, 3, 5\}$.

\begin{figure}[h]
\vspace{-0.5cm}
\centering
\includegraphics[width=3.5in]{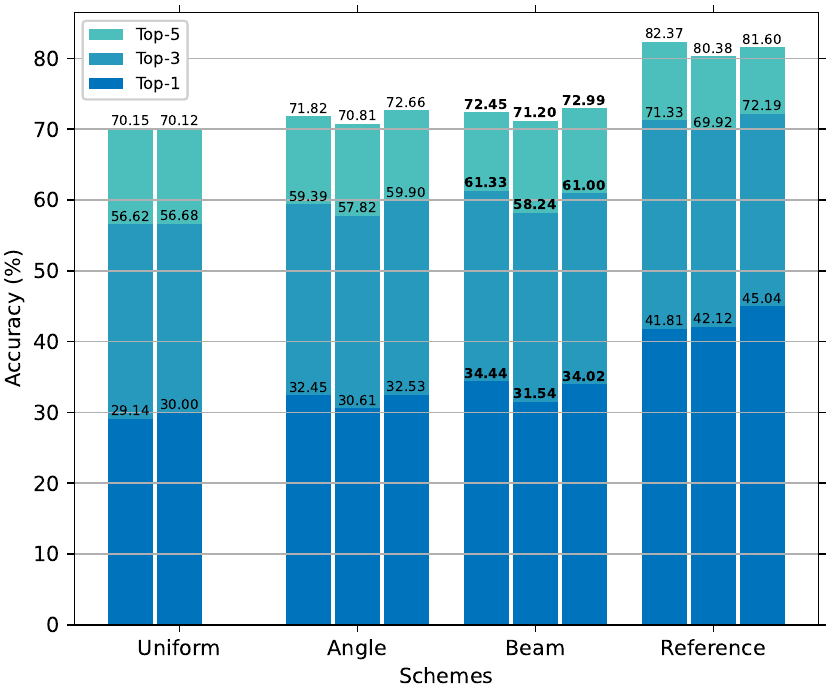}
\caption{Top-$ K $ beam prediction accuracy for different schemes. In each scheme, from left to right are the original scheme, quantization scheme and noise scheme, respectively.}
\label{fig:Top-K}
\end{figure}

%In Fig.~\ref{fig:Top-K}, the detection scheme using uniform beam is not satisfactory, but we believe that it is very suitable for the initialization of trainable beams because compared with random initialization, uniform initialization can avoid the possibility of overlapping beams in the training process to the greatest extent and reduce beam waste. So in this paper, we adopt the uniform codebook to initialize the proposed trainable probe beams.

In Fig.~\ref{fig:Top-K}, the $baseline$ scheme is not satisfactory, but we can initialize the proposed trainable probe beams with the uniform probe codebook. With regard to the remaining schemes, angle\_domain scheme directly uses horizontal and vertical angles $ \{\phi, \theta\} $ as training parameters, while beam\_domain uses $\{u, v\}$ for training. The two schemes are theoretically equivalent, but in practice, it can be found that the scheme with beam-based variables achieves higher accuracy. Besides, we find that the phase quantization of the probe beams seriously degrades the performance, that is, the probe beams are sensitive to the quantization noise. To solve this problem, we propose adding uniform noise on the phases in the training process, and noise adding significantly improves the prediction accuracies on all schemes, including the reference. 

%Finally, the accuracy performance of the beam\_noise scheme is about $13\%$ higher than that of the $baseline$ scheme.

\subsection{Effective Achievable Rate}

\begin{figure}[h]
\vspace{-0.35cm}
\centering
\includegraphics[width=3.5in]{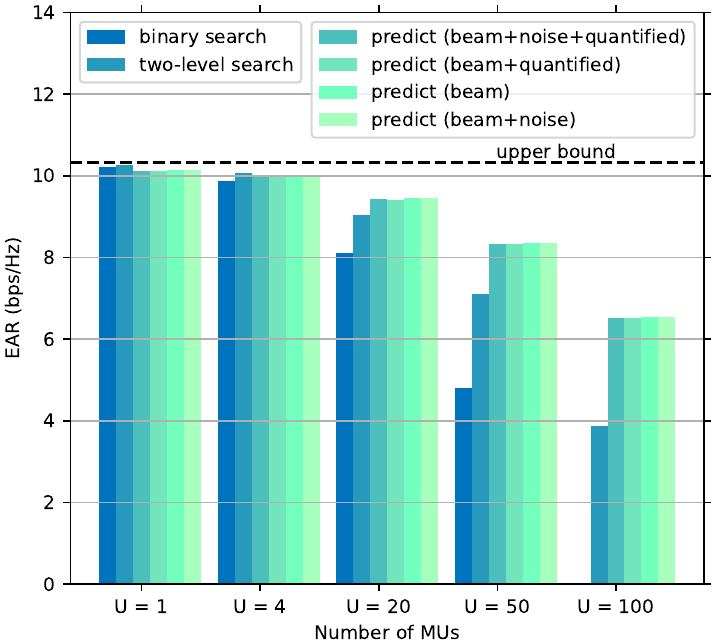}
\caption{EAR versus MU number.}
\label{fig:UE}
\end{figure}

Fig.~\ref{fig:UE} represents the trend of EAR with the number of users at a constant SNR. The two schemes for comparison are the binary search scheme and hierarchical search scheme. Consider a single user, the hierarchical search scheme needs 16 wide beams and 8 narrow beams while the binary search scheme requires $\log_{2}128=8$ rounds of interaction, and the EAR performers of each scheme are close. With the number of users increasing, the performance of the search-based scheme deteriorates rapidly. When the user number reaches 100, the search time of the binary search scheme occupies all the channel coherence time and the corresponding EAR is 0. The hierarchical search scheme spends 60\% of the time on searching for the best beam. In contrast, The prediction-based schemes achieve a 35\% savings in training overhead, and performance remains high.

\section{Conclusion}\label{sec:conclusion}

In this work, we studied the problem of data- and model-driven probe beam training and beam prediction, and utilized deep learning techniques to jointly optimize two networks in an end-to-end way. We also proposed equivalent beamspace variables to train the probe beam module and the noise-adding technology against phase quantization. Simulation results verified the effectiveness of the proposed methods. In future research, we will further extract the channel prior in both frequency and temporal domains via deep learning.

\bibliographystyle{IEEEtran}
\bibliography{reference}

\end{document}